\documentclass{article}
\usepackage{graphicx}
\usepackage{hyperref}
\usepackage[utf8]{inputenc}
\newcommand{\subten}{BRP}
\newcommand{\teniente}{SPL}

\begin{document}
\title{Effectiveness of the GT200 Molecular Detector: A Double-Blind Test}
\author{W. Luis Mochán$^1$ and A. Ramírez-Solís$^2$\\[5pt]
$^1$Instituto de Ciencias Físicas,\\
  Universidad Nacional Autónoma de
  México,\\
  Apartado Postal 48-3, 62251 Cuernavaca, Morelos, México.\\
  \href{mailto:mochan@fis.unam.mx}{\nolinkurl{mochan@fis.unam.mx}}\\[5pt]
$^2$Facultad de Ciencias, Depto. de Fisica\\
  Universidad Autónoma del Estado de Morelos,\\
  Avenida Universidad 1000, 62210 Cuernavaca, Morelos, México.\\
  \href{mailto:alex@uaem.mx}{\nolinkurl{alex@uaem.mx}}\\[5pt]
}
\maketitle 
\begin{abstract}
The GT200 is a device that has been extensively used by the Mexican
armed forces to remotely detect and identify substances such as drugs
and explosives. 
A double blind experiment has been performed to test its
efficacy. In seventeen out of twenty attempts, the GT200 failed in
the 
hands of certified operators to find more than 1600 
amphetamine pills and four bullets hidden in a randomly chosen cardboard
box out of eight identical boxes distributed within a 90m$\times$20m
ballroom. This result is compatible with the 1/8
probability expected for a completely ineffectual device, and is
incompatible with even a moderately effective working one.
\end{abstract}
\section{Introduction}
The GT200 is sold as a {\em remote substance detector} that is claimed by its
manufacturer, UK-based Global Technical Ltd, to detect and identify
various substances including explosives and drugs in tiny quantities,
as small as picograms, and at distances as large as 5km
\cite{folleto}. According to official documents
obtained through the {\em Mexican Federal Access to Information Institute}
(IFAI, for its acronym in Spanish), and from the web
portals of several government agencies \cite{ifai}, the Mexican
Government has bought more than 940 of these devices at prices that
fluctuate from around \$280,000MX to \$580,000MX Mexican pesos (the
current exchange rate fluctuates around \$13MX for \$1USD). Its main
users are the Mexican Army and Navy, with more than 742 and 102 units
respectively, followed by the state petroleum company (PEMEX) with 54
units. According to press releases \cite{prensa}, the GT200 has been
used successfully in hundreds of searches for cocaine, marijuana,
amphetamines, and several other substances. It is not uncommon to find
the GT200 in use at military checkpoints and at airports,  and it has
been used to justify house searches and detentions of an unknown  number
of citizens for the supposedly possession and
trafficking of illegal substances.  Nevertheless, the GT200 is
one of a class of detectors based on dowsing rods with brand names
such as {\em Quadro 
  Tracker, DKL-Lifeguard, Mole, Sniffex, ADE651\ldots} which have
invariably failed in controlled experiments \cite{nij, history}. In
July, 2012, the  manufacturer of the GT200, Gary Bolton, was 
charged in the United Kingdom for dishonestly representing the GT200
as capable of detecting explosives \cite{bolton} and his case is
currently in court. In Mexico,  the
National Committee of Human Rights issued a
recommendation \cite{cndh} against the use of the GT200, as the number
of illegal 
searches by the police and armed forces increased dramatically since
 it was adopted, and its use  would constitute a violation of 
human rights, even if it were functional, as it would violate the
right to privacy.  Currently, the Supreme Court is reviewing the use
of the GT200 to provide {\em evidence}.

One of us (WLM) has participated as expert witness in a judicial trial
where the GT200 
provided multiple evidences of drug and munition possession. As part
of their judicial {\em statements}, the operators of the GT200 stated the {\em
  theory of operation of the apparatus} \cite{teoria}. It is claimed
that  the apparatus is sensitive to {\em diamagnetic and
  paramagnetic fields} that are produced by all substances and which
are characteristic of each, allowing their remote detection and
identification, and that the cards that are used to program the
equipment are fed by electrostatic energy produced by the
operator, among many other statements filled with pseudo-scientific
jargon. WLM was asked to write a detailed criticism of this theory
\cite{critica}, concluding that 
\begin{quote}
  \em it contains numerous conceptual errors and meaningless
  statements that use scientific language but out of context. Those
  statements that do have meaning are false and describe the workings
  of an apparatus that is not compatible with current scientific
  knowledge. Thus, it is certain that the equipment does not work as
  stated by its technical specifications sheet. Furthermore, technical
  arguments yield strong doubts that there is any mechanism whatsoever
  that would allow the device to function and to detect the substances
  that it supposedly detects. The only way to obtain certainty would
  be through a double blind test.
\end{quote}
This study was used in a different judicial trial to free a man that
had been accused of drug trafficking \cite{laura}.
It has also been discussed within the Mexican Senate
\cite{senado,videos}, which 
exhorted the Head of the Executive Branch of the Government of Mexico
to evaluate scientifically the efficacy of the GT200 \cite{exhorto};
the President has not complied yet.

Another one of us (AR) has participated as expert witness in yet another
trial and asked to determine the validity of the evidence provided
through the use of the GT200. In this case the judge ordered the Army
to participate in a scientific test, to be 
conducted by AR, providing a GT200 apparatus, an expert operator and
enough quantity of a substance to be detectable. It is interesting to
note that the Army had previously rejected an offer by Arturo
Menchaca, former president of the Mexican Academy of Science (Academia
Mexicana de Ciencias, AMC) to develop a protocol and supervise a
scientific test. The reason for rejection was the legal obligation to
comply with the terms of the commercial contract signed upon purchase
of the detectors (fragments of the corresponding letters are displayed
in \cite{senado}). AR invited WLM to
participate in the test, which was carried out in October 21, 2011 in
the grounds of the Mexican Academy of Science. Being part of a trial
in progress, we had not made public the results of the
test. Nevertheless, about a year after the test, the main results were
obtained and released to the public by a national newspaper
\cite{laura2}. Thus, we believe there is no longer a reason to keep the
information private. The purpose of this 
article is to describe the test and its results. The paper is
organized as follows. In Sec. \ref{protocolo} we describe the
protocol designed for the test and in Sec. \ref{prueba} the
actual test. We incorporate many details which may be safely
overlooked by the reader, but which may be of interest to those
involved in similar 
tests and which might have some historical
value. The results we obtained are analyzed in Sec. \ref{resultados}, and we
devote Sec. \ref{conclusiones} to our conclusions.

\section{Protocol}\label{protocolo}   

\begin{figure}
  \includegraphics[width=\textwidth]{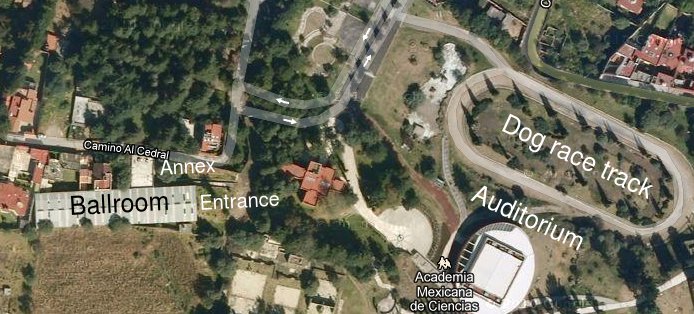}
  \caption{\label{salon}
    Aerial view of part of the grounds of the Mexican Academy
    of Sciences, including the abandoned ballroom where the test took
    place. Its principal entrance and the annex with the secondary
    entrance are shown. As reference to illustrate the scale, the
    auditorium and the old dog race track, now used as a parking lot, are
    also shown. The image is taken from {\em googlemaps}.}
\end{figure}

The test would take place at an abandoned 90m$\times$20m ballroom
(Fig. \ref{salon}) within the grounds of the Mexican Academy of
Sciences. Two members of the army would be designated by the commander
of the 24th military zone in the state of Morelos to participate as
experts in the use, operation and care of the GT200 molecular
detector. They would provide sufficient quantities of any substance that
is expected to be detectable by the GT200. Two experimenters (AR and
WLM) would also participate.

\subsection{Settings}

All the participants would meet at a specified office to identify
themselves and to receive an explanation of the procedures. They would
be divided into a hide ($H$) and a  search ($S$) team, each with one soldier and
one experimenter. All members of each team would stay together
for the duration of the whole test. Up to two witnesses would also be
allowed within each team. Two video-cameras would be positioned within
the ballroom and facing
each other to monitor continuously its two entrances. The
experimenters would also be provided with hand-held cameras to film in
detail the actions 
of the soldier of his corresponding team. To eliminate the
possibility of communication between soldiers during the test, they
would leave all their belongings in bags that would remain at the AMC
offices. Furthermore, the participants would be scanned by a detector
of electromagnetic waves to discard the presence of transmission
devices. The experimenters would keep notes to elaborate a journal in
which the soldiers would sign their agreement. The GT200 would be
operated exclusively by the soldier of the $S$ team. The substance
to be hidden and the boxes which would hide them would be manipulated
exclusively by the soldier of the $H$ team.

Eight opaque cardboard boxes would be used, labeled $AAA$, $AAS$,
$ASA$\ldots\ 
$SSS$. After being examined by the participants, the sample would be
placed within a container, to avoid contaminating  the boxes and the
room. The experimenters would explain to all participants that the
sample would be hidden within one box and that a search for it would
be performed with the detector.

Each box would be inspected by the soldier of the $H$ team to verify
that it is empty and that it is appropriate for the test. Then
he would place each box within the room at the position of his own
choice, making sure that there is enough space between them to allow
detection of the substance if it were in any box, and allowing the
triangulation and all motions required for the successful operation of
the device. Marks would be made in the floor to identify the position
of each box which would not be moved during the rest of the test. The
orientation of the lids of the boxes would be marked and they would
not be modified during the rest of the test. The lids would have to be 
properly closed on their corresponding boxes.  

The soldiers would be asked for objections to the testing procedure
and its conditions; if they were to consider that the conditions
were adequate for the test, it would
proceed. Otherwise, the test would be finished yielding as a result that
the GT200 is unable to detect the sample under the stated
conditions. In any case, the soldiers would sign a document stating
their decision at this point.

\subsection{Experiment}

The test would consist on two iterated series of runs. All
participants would remain silent during each run except for procedural
clarifications. No interruption would be allowed before the ongoing
run were to finish completely. The experimenters would be allowed to
invalidate any run if they consider it violated the protocol, in which
case, it would be repeated and the incident reported.

\subsubsection{First Series: Calibration}

The soldier of the $H$ team would hide the sample in the box labeled
$AAA$. Then, the soldier in the $S$ team would search for it with the GT200,
and write down the identity $xyz$ of the box picked out by the
detector in a specially prepared sheet. The exercises would be repeated for the
remaining boxes $AAS$, $ASA$\ldots $SSS$. The participants would
verify and sign the results sheet.

\subsubsection{Second Series: Test}

\begin{enumerate}
\item The $S$ team would abandon the ballroom and proceed to a waiting room
within an annex (Fig. \ref{salon}). Nobody would remain in the ballroom
but  the members of the $H$ team. 
\item Through a random process, the experimenter $E_H$ of the $H$ team would
  assign one of the eight labels to each of the boxes, and the soldier $S_H$
  would place a paper sheet with the corresponding label {\em inside}
  the designated box and close it firmly.
\item\label{iterate} $E_H$ would toss a coin three times
  and record the three results $xyz$ in a second form. In México, the
  result of a 
  coin toss is either {\em Águila} or {\em Sol}, the equivalent of
  {\em Heads} and {\em Tails} in the US. Thus,  $x,y,z=A$ or $S$  if the
  corresponding result were {\em Águila} or {\em Sol}. Both 
  $S_H$ and $E_H$ would sign the corresponding row of the
  form. The soldier would place the sample within the box labeled
  $xyz$ and after checking that all boxes are firmly closed, the $H$
  team would abandon the ballroom through its main entrance.
\item $E_H$ would knock on the door of the
  waiting room, where after a minute, the $H$ team would enter and
  stay, allowing the $S$ team time to leave the waiting room and to
  enter the ballroom from the annex through a second lateral entrance. Thus, the
  members of both teams would not talk nor see each other. 
\item The soldier $S_S$ of the $S$ team would use the GT200 to identify the
  box containing the sample, and, without opening it, would place an
  indicator on top of its lid. Then, the $S$ team would leave the
  ballroom through its main entrance.
\item The experimenter $E_S$ of the $S$ team would knock on the door of the
  waiting room into which the $S$ team would enter after a minute,
  giving time to the $H$ team time to leave the room and enter the
  ballroom through the lateral entrance.  
\item $S_H$ would identify and record in a third form the label
  corresponding to the box that had been selected by $S_S$ using the
  GT200 as that  containing the sample. The corresponding row would be
  signed by $S_H$ and $E_H$. 
\item $S_H$ and $S_S$ were not to communicate any information about the
  box labeling, the placement of the sample nor the ongoing results of the
  test. 
\item If the run were to be invalidated, a mark would be placed to that effect
  in the three forms.
\item If the test were to be interrupted for any kind of personal
  reasons, it would only be done at this stage.
\item The cycle starting at step \ref{iterate} would be repeated until
  20 valid runs were accumulated.
\item The results would be written out to a fourth form where successes
  and failures would be tallied. All the
  participants would check the transcript for
  consistency with the data contained in the other forms and, when
  satisfied, would write down their name and sign. 
\end{enumerate}
\subsubsection{Statistical Analysis}
The results would be analyzed from a statistical viewpoint to
determine the efficacy of the GT200.

\subsection{Remarks}

The main problems we attempted to solve with this protocol were those
derived from the expected distrust 
among the participants in the test. The possibility of  information
exchange that would give the soldier in the $H$ team knowledge 
about the identity of the box where the soldier in the $S$ team hid
the substance had to be minimized. With eight boxes it would have been enough to
exchange a mere three bits of information. It was also necessary to
minimize the possibility of common excuses in the case of a failure, such as
accusing the 
experimenters  of contaminating the boxes. Furthermore, we expected
the soldiers to distrust the experimenters and it was important to
convince them that no foul play would take place.

The number of boxes was chosen as $8=2^3$, so that three coin tosses
would select a box randomly. The number of trials was chosen as 20, so
that the probability of success in  more than half of the trials would
be less than $10^{-4}$ if the GT200 were as good as chance, as well as the
probability of failure in more than half the trials were the detector
moderately efficient, with a probability of success of 85\% in an
individual run. Thus, success or failure would be definite.

The purpose of the calibration series, which we expected to be
successful,  was to eliminate excuses were the test to fail,
as the conditions during both series of detection trials would be
essentially identical.  

\section{Test}       \label{prueba}       
The test took place on Oct. 21, 2011 at the premises of the Mexican
Academy of Sciences. Two Ministerial Policemen were in custody of part
of the sample consisting of 1630 30mg capsules of {\em Itrabil} and
33 30mg capsules of {\em Obeclox}, both with the substance {\em
  Clorobenzorex}, a stimulant drug of the phenethylamine and
amphetamine chemical families. The rest of the sample consisted of 3
9mm$\times$19mm bullets and one 0.28'' bullet, whose custody was in
charge of two soldiers, a Lieutenant Colonel and a Sergeant 2nd
Class. The sample had purportedly been detected using the GT200 and
later confiscated from the house of the defendant.

The operators of the GT200 were the Lieutenant Saulo
Pérez-Lozano (SPL) and the Sublieutenant Bernabé Reyes-Pérez
(BRP). The Lawyer María Elena Gómez-Salgado (MEGS) and the Sergeant
2nd Class Jonathan Juárez-Ibarra (JJI) witnessed the test, which
was conducted by AR and WLM. Two fixed video-cameras were
set up in opposite sides of the room and filmed the whole test.  Two
hand-held cameras were operated by the experimenters and filmed parts
of the test \cite{video}.

The operators of the GT200 were judged to be fully confident
in the capabilities of the device before the test began. Thus, the
experimenters judged subjectively that they would not attempt to
cheat, and they were not  scanned for electronic devices
after they were asked to leave behind their belongings.
A simple inspection of their pockets revealed they were empty.

After receiving an explanation of the procedures, the ballroom where
the test would be conducted was examined. At first, SPL worried about 
the neighboring houses; he wanted to rule out the possibility of their
occupants being sick and taking medicine, as that could confuse the
detector, which might link energetically with a substance outside the
premises. Nevertheless, he decided the room 
allowed the boxes to be placed far enough from the walls to minimize
the effects of neighboring houses. On the other hands, he asked that a
table with coffee and soft drinks be removed from the room. He also
complained that the manila envelope that contained the bag of
amphetamines smelled of marijuana, and thus, it would have
contaminated the exterior of the bag, which in turn would contaminate the
interior of the boxes. Although marijuana was not among the substances
to detect, and the ministerial policemen stated that the envelope was
brand new and denied that it had never been in contact with marijuana
or any similar substance, it was decided to place 
the sealed bags with the pills and the bullets 
within a clean clear plastic bag and to close it tightly. The
experimenters convinced SPL that 
although the interior of the bag might get contaminated, its exterior
would remain clean and unable to contaminate the boxes. Thus, he
agreed that the test could proceed. 
\begin{figure}
  \includegraphics[scale=.65]{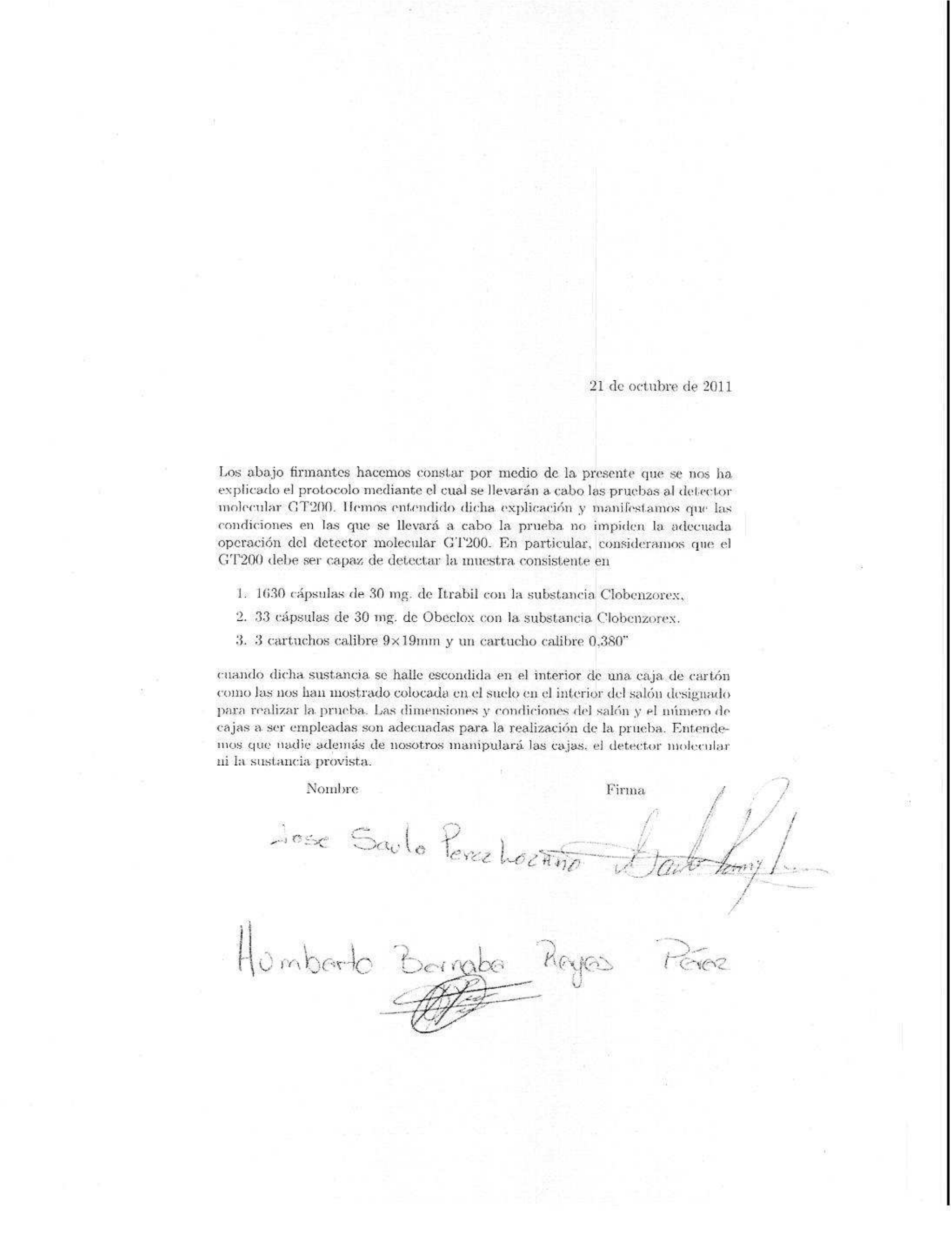}
  \caption{\label{agree}Agreement on the conditions of the test.}
\end{figure}
Figure \ref{agree} shows the agreement signed by the soldiers, which
translated reads
\begin{quote}
  \em The undersigned certify that we have received an explanation of
  the protocol through which the GT200 will be tested. We have
  understood the explanation and we state that the conditions under which
  the test will take place don't hinder the adequate operation of the
  GT200 molecular detector. In particular, we believe the GT200 should
  be capable of detecting the sample, consisting of
  \begin{enumerate}
  \item 1630 capsules of Itrabil 30mg containing Clorobenzorex,
  \item 33 capsules of Obeclox 30mg containing Clorobenzorex,
  \item 3 9$\times$19mm bullets and one 0.380'' bullet,
  \end{enumerate}
when hidden within a cardboard box such as those we have been shown and
placed within the room where the test will be held. The size and
conditions of the room, as well as the number of boxes to be employed,
are adequate for the test. We understand that nobody but us is to manipulate
the boxes, the molecular detector nor the sample.
\end{quote}

The soldiers decided that the GT200 would be operated by BRP. Thus, 
the members of the $H$ team were AR as experimenter and SPL as
soldier, and the members of the $S$ team were WLM as experimenter and
BRP as operator of the GT200. The $S$ team was joined by MEGS and JJI
as witnesses. The policemen and the lieutenant colonel that were in
charge of the sample remained out of the ballroom and were asked not
to talk to the team members while the test was in progress.

\begin{figure}
  \includegraphics{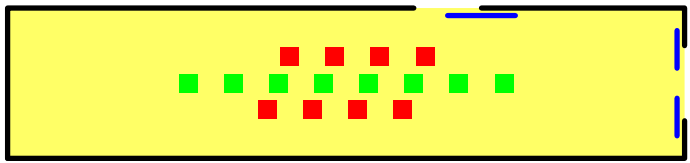}
  \caption{\label{place}Schematic drawing of the ballroom, indicating
    the entrances and the position chosen for the boxes in the initial
    (green) and the last (red) part of the test.}
\end{figure}

SPL was asked to place the boxes at the positions of his choice. He
decided to place them along a straight line oriented from east-west,
along the long side of the ballroom, at a distance of approximately 6m
from each other and at a distance of 10m from the north and south
walls (Figs. \ref{salon} and \ref{place}). 

\begin{figure}
  \includegraphics[scale=0.65]{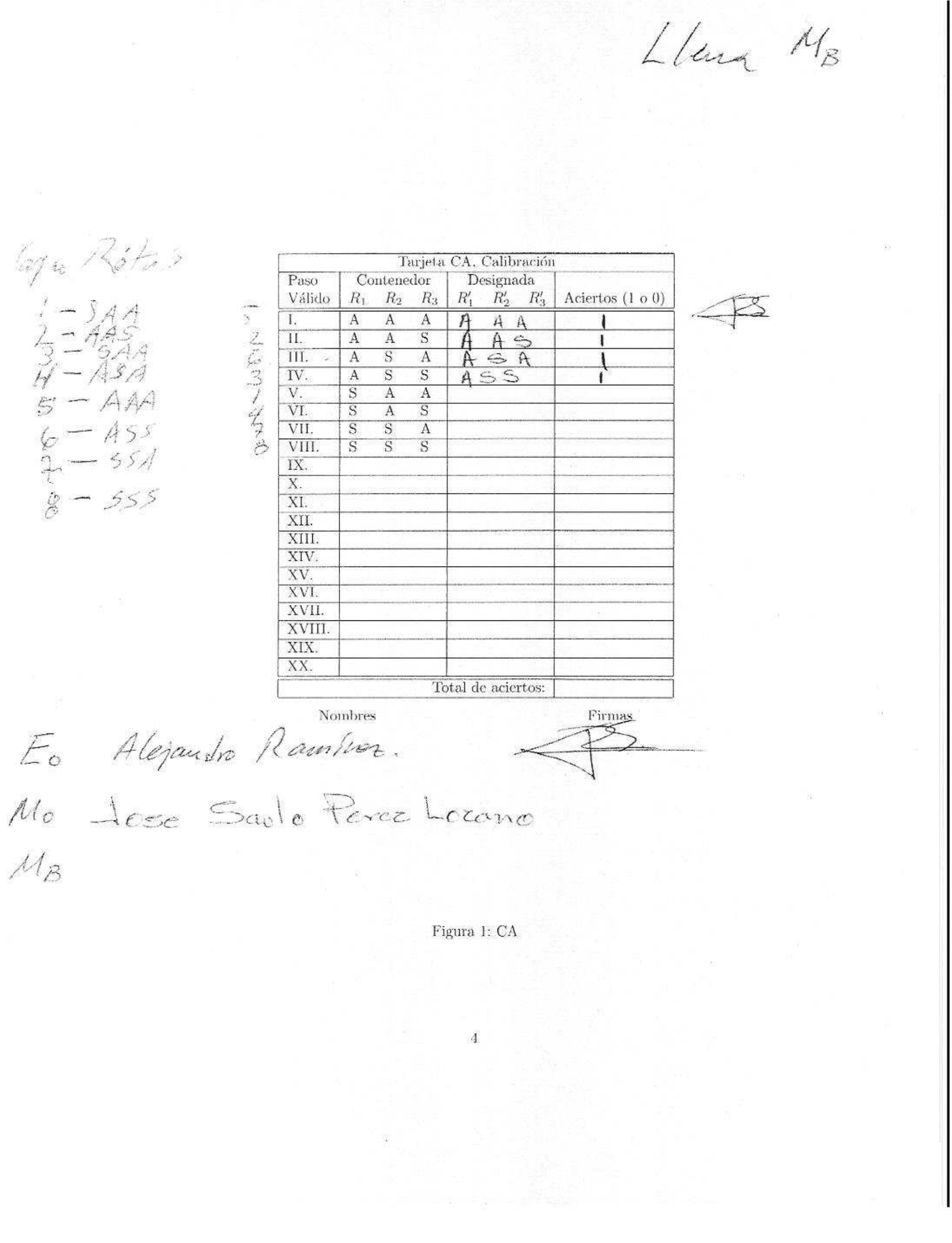}
  \caption{\label{calib}Results of the calibration stage of the
    test. }
  
\end{figure}

The calibration stage began at noon. SPL hid the sample in box
$AAA$ (the rightmost in Fig. \ref{place}) in plain view of all
the participants. BRP searched it with the GT200 and after pacing 
a couple times the length of the area occupied by the boxes, the
antenna of his GT200 repeatedly rotated 90$^\circ$ in front of box
$AAA$, so that, after 5 minutes, he announced that the sample
was located indeed in box $AAA$. This result was written down in the
appropriate sheet (Fig. \ref{calib}).
The test proceeded successfully with box $AAS$, for which finding the
sample took 4 minutes. Successful detection in box $ASA$ took 5 minutes and after
finishing, BRP said he couldn't go on. He told us that he was becoming
very tired and that the GT200 stops working if its operator is
fatigued. Not wanting to allow excuses for a possible failure, at the
end of that run we 
interrupted the test and took a 7 minute break. After searching
successfully for the sample in box $ASS$, a task that took another 7
minutes, BRP said it would be impossible to finish the test as
planned. Considering that the second stage would be more important,
and that the first stage had enough repetitions for its intended
purpose, we decided to conclude at this point the calibration
stage. As shown in Fig. \ref{calib}, the GT200 was able to correctly
locate the sample in four out of four attempts. At this point the test
was interrupted and we took a 10 minute break.

\begin{figure}
  \includegraphics[scale=0.58]{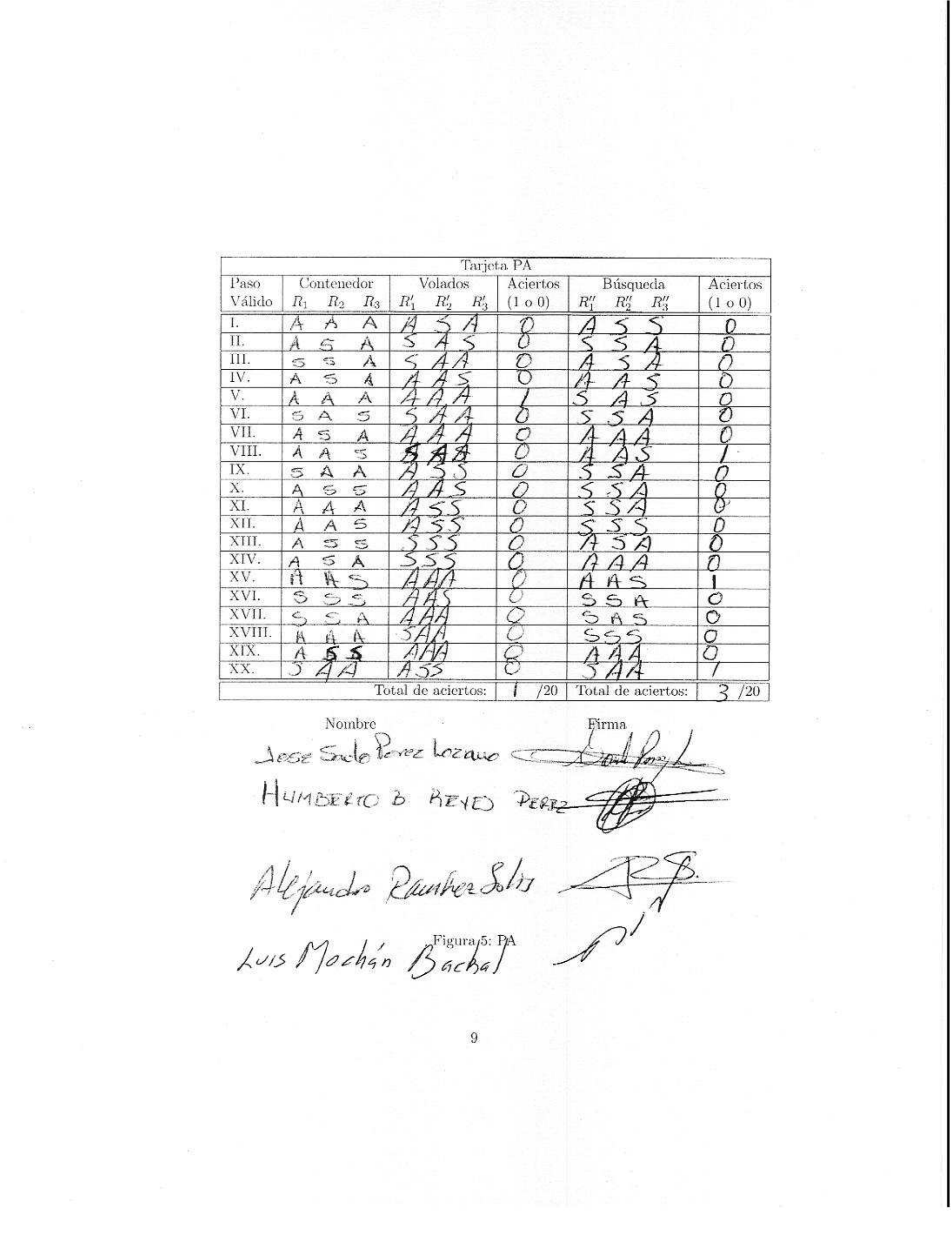}
  \caption{\label{results}Summary of the results. From left to right,
    the columns indicate the consecutive search number, the label of
    the box where the sample was actually hidden, a label chosen
    through three flips of a coin, number of correct hits, the label
    of the box picked out through a search with the GT200 and the
    number of correct hits.} 
\end{figure}

The  {\em double blind} stage began at 12:41. The $S$ team entered the
waiting room while the $H$ team entered the ballroom and drew randomly from a
recipient a sequence of 
folded papers marked with the box labels to be assigned  to the
boxes. Thus, the first box counting from the entrance at the 
right towards the left in Fig. \ref{place} was assigned the label
$SAA$, the second $AAS$, the next boxes  $ASS$, $SAS$, $AAA$, $ASA$,
$SSA$, and finally, the eighth box was marked $SSS$ (see table
\ref{cajas}).
\begin{table}
  $$
  \begin{tabular}{r|rrrr}
    Box&\multicolumn{4}{c}{Labels}\\
    Number&Phase 1&Phase 2&Phase 3&Phase 4\\
    \hline
    \hline
    1&SAA&SSS&SAS&AAS\\
    2&AAS&ASS&SSA&ASA\\
    3&ASS&SAA&AAS&ASS\\
    4&SAS&SSA&SSS&SAA\\
    5&AAA&ASA&SAA&SSA\\
    6&ASA&AAS&ASA&SSS\\
    7&SSA&AAA&AAA&SAS\\
    8&SSS&SAS&ASS&AAA\\
  \end{tabular}
  $$
  \caption{\label{cajas}Assignment of labels to boxes during the four
    phases of the double blind stage. The boxes were numbered from
    East to West (left to right in Fig. \ref{place}}
\end{table}
 SPL put each label 
inside of the corresponding box,  so that they were not visible with
the lid on. AR
tossed a coin three times, obtaining  {\em  Águila} each
time. Therefore, SPL put the sample in the box corresponding to 
the label $AAA$, the fifth counting from the entrance. The $H$ team
left the room and AR signaled the $S$ team to start the search. After
BRP picked out a box with his GT200, the $S$ team left the room, the
$H$ team entered, identified the box supposedly containing the hidden substance
 and wrote down its label. The
search took 11 minutes. Subsequent searches took 12, 6, 4 and 19
minutes. The results are summarized in fig. \ref{results}. While
timing the searches was not considered in the protocol, their
fluctuations, going from 4 up to 24 minutes are interesting, and they
are summarized in Table \ref{tiempos}. (Unfortunately, the times
corresponding to two 
of the three successful searches were not registered: after search
XV due to distractions arising from the end of a stage and a
discussion with the operator, and at the end of search XX due to
the end of the test.)    
\begin{table}
  $$
  \begin{tabular}{rrrr}
    Number of&Starting&Ending&Duration\\
    Search&Time&Time&(minutes)\\
    \hline
    \hline
    I&12:45&12:56&11\\
    II&13:18&13:30&12\\
    III&13:38&13:44&6\\
    IV&13:53&13:57&4\\
    V&14:07&14:26&19\\
    Lunch, swap, reshuffle\\
    VI&14:50&15:14&24\\
    VII&15:20&15:28&8\\
    VIII&15:36&15:44&8\\
    Rest\\
    IX&15:59&16:06&7\\
    X&16:12&16:25&13\\
    Swap, reshuffle\\
    XI&16:40&16:46&6\\
    XII&16:55&17:05&10\\
    XIII&17:14&17:21&7\\
    XIV&17:29&17:39&10\\
    XV&&&\\
    Swap, reshuffle\\
    XVI&18:03&18:11&7\\
    XVII&18:17&18:25&8\\
    XVIII&18:31&18:38&7\\
    XIX&18:44&18:48&4\\
    XX&18:56&&\\
  \end{tabular}
  $$
  \caption{\label{tiempos}Times taken for each search during the
    double-blind stage. 
}
\end{table}
For comparison purposes, we also asked the soldier of the $S$ team to
flip a coin three times after each search and record the result, which
was later transcribed into the results sheet,
Fig. \ref{resultados}. By the end of the fifth iteration we took a
lunch break. 

During lunch, \subten\ complained that he was getting extremely tired
and suggested 
that the test  be extended for a few days. AR and WLM
decided that, to comply with the judge's orders, the test would have to
be finished in only one day. To allow \subten\ some rest, WLM asked
\teniente\ if he would be able to operate the GT200. As it turned out
that he was an even more experienced operator than \subten, it was
decided to swap operators between the $S$ and $H$ teams at the end of each
fifth run. Unfortunately, swapping operators
allowed them to know the ongoing results of the
test before it was completely finished, something that AR and WLM had
wanted to avoid in order to comply fully with the double-blind
character of the test. Furthermore, \teniente\ decided that the
disposition of the boxes should be changed to reduce the distance to
walk during each search. AR and WLM agreed provided
the new disposition would not hinder the efficacy of the
detection. The new disposition was in a zig-zag pattern with
approximately $6m$ between neighboring boxes, as shown in
Fig. \ref{place}.

The labels were reshuffled (Table \ref{cajas}) and the second phase of
the double blind stage
began, with \teniente\ in the $S$ team and \subten\ in the $H$
team. The test was resumed at 14:50 and the sixth search took 24
minutes. \teniente\, being aware of the failure of the first five
searches (Fig. \ref{resultados}), complained that the boxes might have
been already contaminated by marijuana vapors from the manila
envelope. WLM insisted this couldn't happen as the sample had been
placed within a closed plastic bag. Furthermore, marijuana vapors shouldn't
interfere with the test, as it was not the substance that was being
searched for. It turned out that \teniente\ had actually placed a card for
marijuana within his GT200, besides one for amphetamines and another
for munitions. He explained that the extra card enhanced the detecting
power of the GT200. At the end of the seventh and eight iterations,
\teniente\ was sweating abundantly and he complained 
that the waiting room was one floor below the ballroom and that
climbing steps left him agitated, hampering the efficacy of the
search. WLM told him he could rest in the ballroom as long 
as he desired before starting each search and announced a fifteen
min. break. After two more searches, taking 7 and 13 minutes
respectively, the second phase was finished.

The pattern was repeated for two more phases, swapping of operators,
reshuffling of box labels and five hide and search iterations. 
At the
end of the third phase \subten\ complained that to be effective, the
GT200 instruction manual indicates that complementary search methods
should be
used, such as a trained companion dog or a colleague to perform an exhaustive
manual search. These arguments were dismissed by WLM, as it was the GT200,
not the dog or a companion that was being tested. Furthermore,
\subten\ believed that the GT200 was frequently picking up boxes that were
contiguous to those actually containing the sample, and pointed out that the
GT200 is not expected expected to point at the exact location of the
sample, but only to the general area. WLM argued that the instruction
manual, which he read during the test, stated that the uncertainty was
of only 2m, which should have been good enough to identify the box
containing the sample, as the boxes had been placed by the soldiers at the
positions of their choice and with a large enough distance, at least 6m,
between each other.

\section{Results}\label{resultados}
As shown by Fig. \ref{calib}, the GT200 pointed out to the correct
location during the calibration stage in four times out of four
attempts. This demonstrates that the GT200 was perfectly capable of
finding the sample {\em when the operator knew beforehand the location
  where it had been hidden}. The probability of obtaining this result
by chance only would have been $(1/8)^4\approx 0.00024$, rather
small. Thus, this stage also showed that the experimental conditions
were adequate for the search of the provided sample and that detection
was not being hindered by the kind of boxes employed, the envelopes
and bags that contained the sample, the neighboring homes, the
presence of the researchers and witnesses, the positioning of the
boxes within the room, the walls of the room, the weather conditions
nor by any other condition present during the test.

\begin{figure}
  \includegraphics{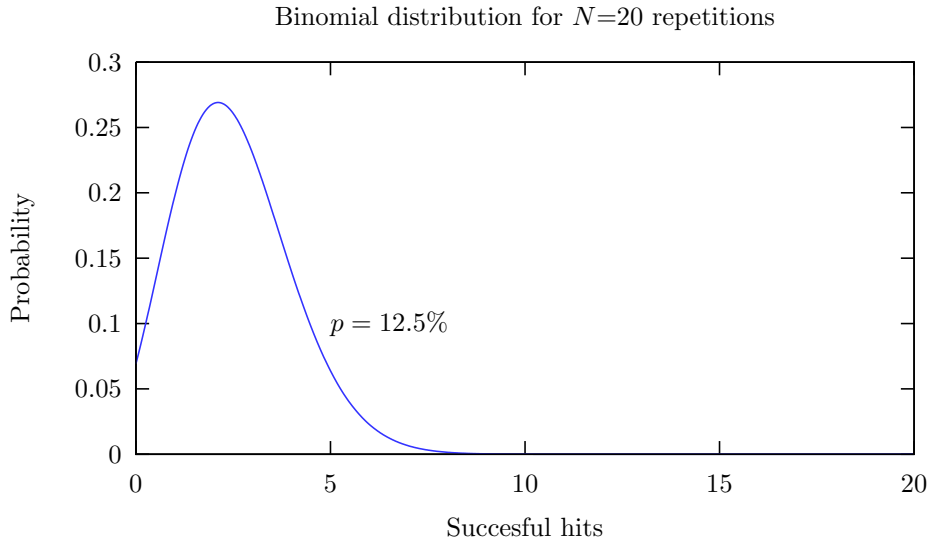}
  \caption{\label{binomialeng}Binomial distribution showing the
    probability of different experimental outcomes for a completely
    random detector.} 
\end{figure}

\begin{figure}
  \includegraphics{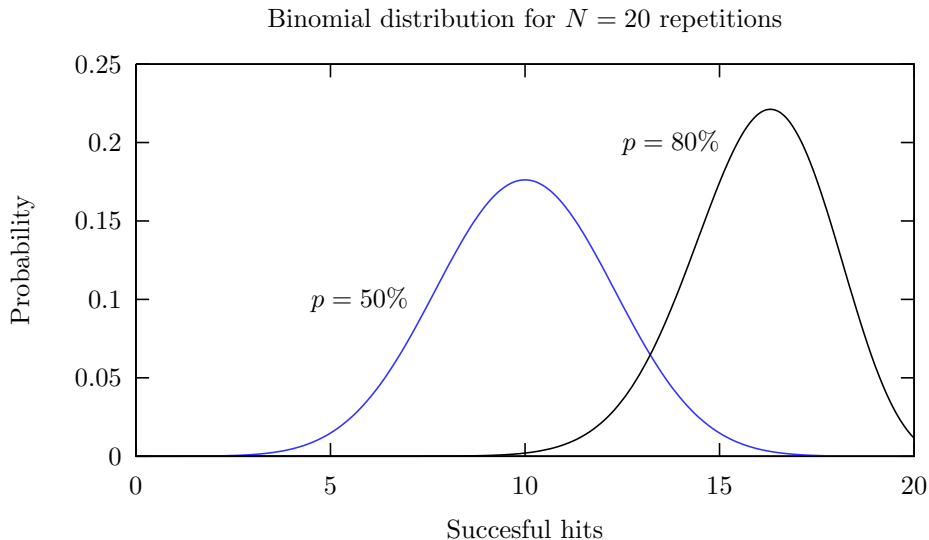}
  \caption{\label{binomialchaf}Binomial distribution showing the
    probability of different experimental outcomes for moderately
    effective detectors.}
\end{figure}

As shown by Fig. \ref{results}, during the double-blind stage, the
operators of the GT200 were able to 
identify the correct box containing the sample in only three out of
twenty attempts. This result is consistent with the binomial
distribution 
\begin{equation}
  \label{binomial}
  b(p,m,N)=\frac{N!}{m! (N-m)!} p^m (1-p)^{N-m}
\end{equation}
that describes the probability of having $m$ correct hits out of $N$
attempts when the efficacy of the detector, i.e., its probability of
success in any one run, is 
$p$. For a completely random case, the probability of a successful
search when the sample is hidden in one out of eight boxes would be
$p=1/8$. The corresponding binomial distribution is illustrated in 
Fig. \ref{binomialeng}.
It has an average of 2.5 and a standard deviation of 1.48, and
therefore is completely consistent with the experimental result.
The figure shows that in a random search the probability of having
obtained $m\ge7$ hits would be negligible, more than three standard
deviations away from the mean. In fact, the probability of less
than 7 hits would be above 99\%. The probability of having exactly 3 hits is
$b(1/8,3,20)=23\%$, very close to the maximum value. On the other
hand, in Fig. \ref{binomialchaf}
we show the corresponding distribution function corresponding to
a moderately effective detector with an efficacy of $p=80\%$. This
distribution would yield an average of 18 successful hits with a 
standard deviation of 1.8. Our experimental result is completely
inconsistent with even this moderate efficacy, being more than 8
standard deviations away from the mean, with a probability of less
than $10^{-9}$. Even if the efficacy of the detector were as low as
$p=50\%$ (Fig. \ref{binomialchaf}), the probability of having obtained
only three successful hits would be a mere $b(0.5,3,20)=0.1\%$. 

\begin{figure}
  \includegraphics{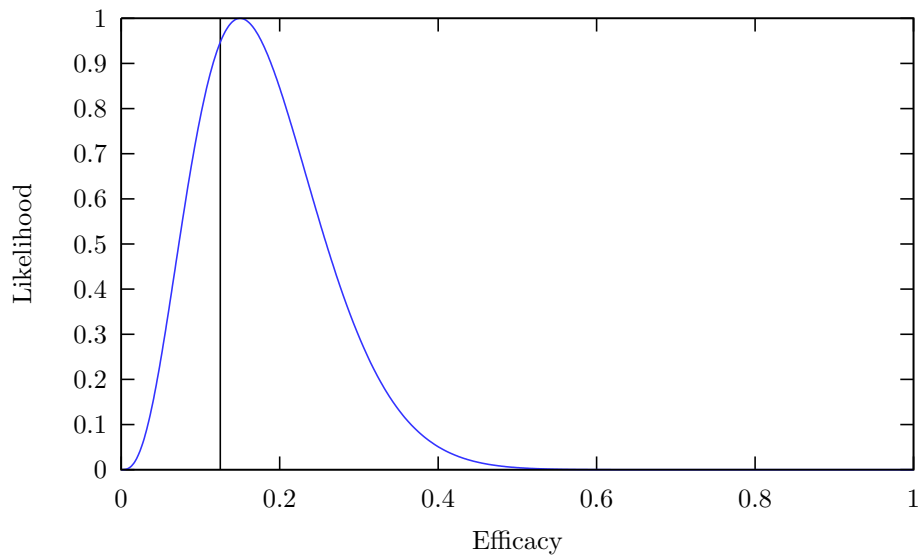}
  \caption{\label{likelihood}Likelihood of the detector having an
    efficacy $p$ given the experimental result of 3 successful hits out
    of 20 tries. The vertical line corresponds to complete randomness,
  $p=1/8$. }
\end{figure}

\begin{figure}
  \includegraphics{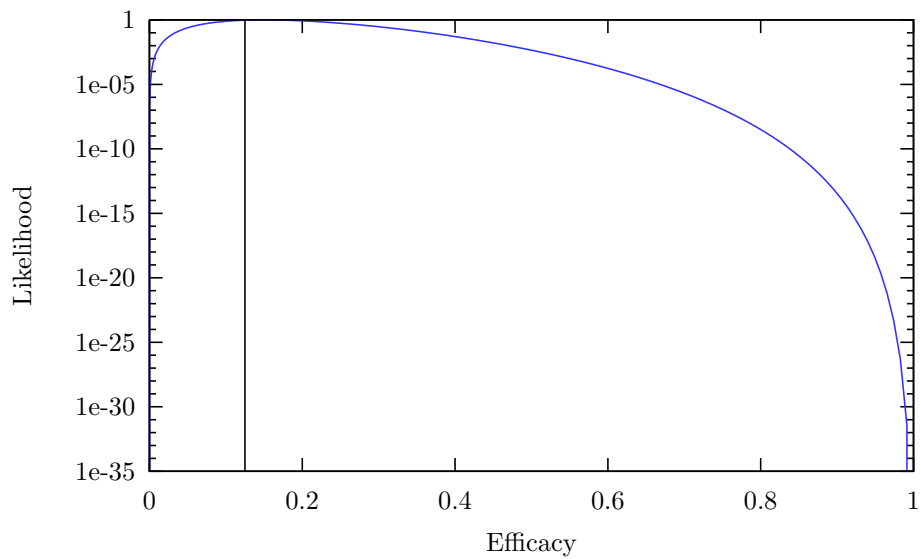}
  \caption{\label{loglikelihood}Likelihood of the detector having an
    efficacy $p$, as in Fig. \ref{likelihood}, but displayed with a
    logarithmic vertical scale.}
\end{figure}

The results may be
summarized by Fig. \ref{likelihood}
which displays the likelihood $b(p,3,20)$ that the detector has an
efficacy $p$ normalized to its maximum value $b(3/20,3,20)$. The
figure shows that 
the normalized likelihood of a completely ineffective detector, behaving in a
completely random fashion with $p=1/8$ is quite high, $0.95$, while
the likelihood of any value $p\ge0.5$ is negligible, less than
$0.005$, and a rapidly decreasing function of $p$. This is made
evident in Fig. \ref{loglikelihood}, where the likelihood is plotted
in a logarithmic scale.

To compare the actual search results with those expected from a fully
random search, the operators had been asked to flip a coins three
times and register the corresponding results in a specific column
within the table of results (Fig. \ref{results}). Unfortunately, the
result of the coin flips (only one successful hit out of twenty tries) was worse
than that of the detector, so that, although perfectly compatible with
chance (we would expect this result in one out of five repetitions of
the complete test, as shown by Fig. \ref{binomialeng}), its comparison
to the detector performance lacked impressiveness.

\section{Conclusions}\label{conclusiones}
The test described in this paper allows us to conclude that the GT200
proved completely {\em ineffective} as an instrument to detect the
substances and munitions used as the sample when the operator ignores
beforehand where the substance has been hidden. It is important to
note that its manufacturer and its users have claimed that the GT200
detects and identifies  nanograms, and even picograms, of hundreds of
substances such as 
manifold drugs and explosives from distances as far away as hundreds or even
thousands of meters while hidden in presumably unknown locations,
while, in this test, we used more than a kilogram of pills containing
about 50g of the stimulant drug Clorobenzorex, as well as four
bullets. They were known to be hidden in one out of eight boxes, and
the operators were free to explore from distances not larger than 100m
and to approach the boxes as close by as desired. It is also important to remark
that the sample employed in this test had purportedly been previously
detected within a house from its outside and forms part of the {\em
  evidence}  being currently
employed against its dweller, accused of illegal drug dealing. During
this test,  
the GT200 picked consistently
the correct location of the sample only when its operators knew
beforehand where it was hidden; when they didn't, {\em the GT200
  failed absolutely and threw results fully consistent with a random
  choice}. Thus,  the GT200 is necessarily 
manipulated by its user to point towards the location where he
expects the sample to be hidden, although he may be unaware
of this manipulation.  The GT200 itself provides no information
about the location of the sample, even when used by
trained and certified operators. Thus, {\em we conclude that the
  GT200 is worthless as a substance detector.}

\section*{Acknowledgments}
We are grateful to the Mexican Academy of
Sciences and in particular, to its former president, Dr. Arturo
Menchaca,  and its Executive Secretary, Dr. Renata Villalba-Cohen, for
allowing us the use of their installations and for their support
before and during the test. This work was partially supported by
DGAPA-UNAM under grants. No. IN20909 and IN108413 (WLM).

\end{document}